\documentclass[12 pt]{amsart}
\usepackage{graphicx}\usepackage [latin1]{inputenc}
 \theoremstyle{definition}
 
 \theoremstyle{remark}

 \numberwithin{equation}{section}

\begin{document}
\title[Violating the bilocal inequality with separable mixed states]{Violating the bilocal inequality with separable mixed states in the entanglement-swapping network}

\author{Shuyuan Yang, Kan He$^*$}
\address[Shuyuan Yang]{College of Information and Computer \& College of Mathematics, Taiyuan University of Technology, Taiyuan,
030024, P.R. China}

 \email{yangshuyuan2000@163.com}

\address[Kan He$^*$]{Key Laboratory of Quantum Information,
University of Science and Technology of China, Chinese Academy of
Sciences, Hefei, 230026, P. R. China}

 \email{hekanquantum@163.com}

\thanks{ $^{*}$Corresponding author}
%\thanks{{\it Primary number} 47B48,47B49.}
\thanks{{\it Key words and phrases:} Bilocal inequality, Werner state, Particle swarm optimization (PSO) algorithm}
%\thanks{The work is supported by National Natural Science Foundation of China under Grant No.11771011, 12071336.}

\begin{abstract}
It has been showed that two entangle pure states can violate the
bilocal inequality in the entanglement-swapping network, vice versa.
What happens for mixed states? Whether or not are there separable
mixed states violating the bilocal inequality? In the work, we
devote to finding the mixed Werner states which violate the bilocal
inequality by Particle Swarm Optimization (PSO) algorithms. Finally,
we shows that there are pairs of states, where one is separable and
the other is entangled,  can violate the bilocal inequality.

\end{abstract}
\maketitle

\section{Introduction and preliminaries}

Quantum nonlocality, detected by violation of Bell inequalities, is
a significant property of quantum mechanics \cite{JS}. Nowadays,
it has become a key tool in the modern development of quantum
information and its applications cover a variety of areas
\cite{NDSV}: quantum cryptography \cite{ANNSS}; complexity theory
\cite{MAH}; communication complexity \cite{HRS}; Hilbert space
dimension estimates \cite{JHB}.

Recently, generalizations of Bell's theorem for more complex causal
structures have attracted growing attention
\cite{CNS1,CDNS,TF,APDA,RRJD,CB,RCD,FS,TF2}. Some researchers have
studied the characteristics of classical, quantum and post-quantum
correlations in networks by constructing the network Bell
inequalities and exploring their quantum
violations \cite{ML,ERT,EAMDAM,MESNNS,MYSSNN,NJYAESN}. In this paper,
we focus on the simplest nontrivial network Bell experiment, known
as the bilocality scenario, with only three observers and two
sources. It features two independent sources that each produce a
pair of particles. The first pair $S_1$ is shared between observers Alice
and Bob while the second pair $S_2$ is shared between Bob and another
observer, Charles (see Fig.1). Consider that Alice receives
measurement setting (or input) $x$, while Bob performs measurement
$y$, and Charles $z$. After measuring the three parts, they obtain
outcomes denoted $a$, $b$ and $c$, respectively. Bob's
measurement $y$ might correspond to a joint measurement on the two
systems that he receives from each source. The correlations between
the measurement outcomes of the three parties are described by the
joint probability distribution $p(a,b,c|x,y,z)$.

\centerline{\includegraphics[width=4.5in]{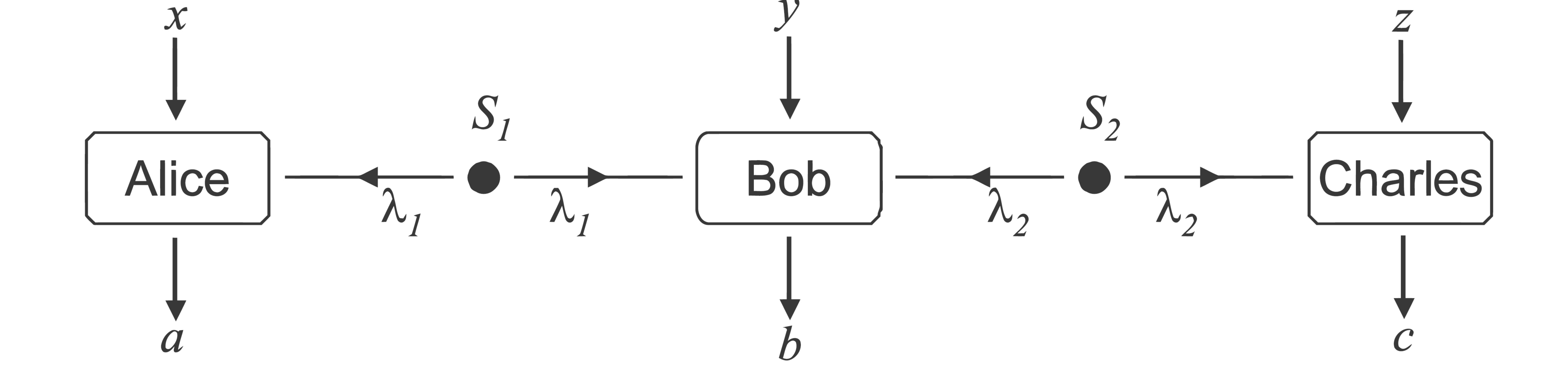}}
%\vspace*{6mm}

\vskip 1mm

%\centerline{\footnotesize Í¼1\ \ \  Gd$_{60}$Fe$_{30}$Al$_{10}$ Ìõ´øµÄ´Å»¯ÇúÏß \quad (origin µÈÈí¼þ»æÖÆµÄ²åÍ¼)}

%\vskip 1mm

\centerline{\parbox[c]{5cm}{\footnotesize Fig.~1.~Scenario of bilocality}}

\vskip 0.55\baselineskip

In our scenario, two independent sources distribute mixed states, $\rho_{AB}$ and $\rho_{BC}$, between three distant
observers, Alice, Bob and Charles. The tripartite joint probability distribution $p(a,b,c|x,y,z)$ is $2$-local if it can be written as
\begin{equation*}
p(a,b,c|x,y,z)=\int\int d\lambda_1 d\lambda_2 q_1(\lambda_1)q_2(\lambda_2)p(a|x,\lambda_1)p(b|y,\lambda_1,\lambda_2)p(c|z,\lambda_2),
\end{equation*}
where $\lambda_1$ and $\lambda_2$ are the independent shared random variables distributed according to the densities $q_1(\lambda_1)$ and
$q_2(\lambda_2)$, respectively.

It has been documented that the set of Bell-local (or 1-local) correlations can be fully characterized by
linear Bell inequalities \cite{NDSVS}. So how do we represent bilocality correlation sets?
Fortunately, the nonlinear inequality in Ref \cite{CDNS} can be used to efficiently capture bilocality correlations. This nonlinear inequality is called bilocality inequality, which is satisfied by bilocality correlations but can be violated by non-bilocal correlations. Consider that Alice
and Charles receive binary inputs, $x=0,1$ and $z=0,1$
and give binary outputs, denoted $a_x=\pm1$ and
$c_z=\pm1$, respectively. The middle party Bob always
performs joint measurement, denote
Bob's outcome by two bits $b_0=\pm1$ and $b_1= \pm1$. Then,
the bilocality inequality can be written as:
\begin{align} \label{1.1}
S\equiv\sqrt{|I|}+\sqrt{|J|}\leq 2,
\end{align}
where
\begin{equation}\label{1.2}
I\equiv\sum_{x,z}\langle a_xb_0c_z\rangle=\langle(a_0+a_1)b_0(c_0+c_1)\rangle,
\end{equation}
\begin{equation}\label{1.3}
J\equiv\sum_{x,z}(-1)^{x+z}\langle a_xb_1c_z\rangle=\langle(a_0-a_1)b_1(c_0-c_1)\rangle.
\end{equation}
The bracket $\langle\cdot\rangle$ denotes the expectation value of many experimental runs.

Gisin et al \cite{GMQTA} showed that  pairs of pure states can
violate bilocality inequality  if and only if the two pure states
are entangled. Naturally, we ask the question: what kind of mixed
states can violate the bilocality inequality? Whether or not are
there separable mixed states violating the bilocal inequality? In
the work, we devote to finding the mixed Werner states which violate
the bilocal inequality by Particle Swarm Optimization (PSO)
algorithms. Finally, we shows that there are pairs of states, where
one is separable and the other is entangled,  can violate the
bilocal inequality.

\section{Arbitrary pairs of Werner state}

Let us first consider that Bob performs measurement $b_y$ has trace zero. That is, $b_0=\sum_{ij}m_{ij}\sigma_i\otimes \sigma_j$, $b_1=\sum_{ij}n_{ij}\sigma_i\otimes \sigma_j$ (with $i,j\in\{1,2,3\}$) and eigenvalues of $b_{0(1)}$ have to lie in $[-1,1]$.
$\sigma_i$ denotes the Pauli matrices. Then, $a_i$ and $c_i$, $i=0,1$, are Hermitian operators with eigenvalues $\in[-1,1]$. In particular, they can be expressed as: $a_{i}=x_{i1}\sigma_1+x_{i2}\sigma_2+x_{i3}\sigma_3$ and $c_{i}=y_{i1}\sigma_1+y_{i2}\sigma_2+y_{i3}\sigma_3$, where $\overrightarrow{x_i}=(x_{i1},x_{i2},x_{i3})$ and $\overrightarrow{y_i}=(y_{i1},y_{i2},y_{i3})$ are four unit-length vectors.

Every bipartite state involving two qubits can be written in
the form
\begin{equation}\label{2.1}
\rho=1/4(I\otimes I+
\mathbf{r}\cdot\overrightarrow{\sigma}\otimes I+I\otimes\mathbf{s}\cdot\overrightarrow{\sigma}+\sum_{m,n=1}^3t_{mn}^{AB}\sigma_m\otimes \sigma_n)
\end{equation}
where, in the notation of \cite{RPM}, $I$ is the $2\times2$ identity operator, ${\{\sigma_n\}}^3_{n=1}$ are Pauli matrices, $\mathbf{r}$ and $\mathbf{s}$ are Bloch vectors in $\mathbb{R}^3$, and $\mathbf{r}\cdot\overrightarrow{\sigma}=\sum^3_{i=1}r_i\sigma_i$, $t_{mn}=Tr(\rho\sigma_n\otimes\sigma_m)$ forms a matrix denoted $T_\rho$.

Let
\begin{equation}\label{2.2}
\rho_{AB}=1/4(I\otimes I+
\overrightarrow{m_A}\cdot\overrightarrow{\sigma}\otimes I+I\otimes\overrightarrow{m_B}\cdot\overrightarrow{\sigma}+\sum_{i,j=1}^3t_{ij}^{AB}\sigma_i\otimes \sigma_j)
\end{equation}
be the state shared by Alice and Bob.  Similarly
we express $\rho_{BC}$, the state shared by Bob and Charles, which can be written in the form $\rho_{BC}=1/4(I\otimes I+
\overrightarrow{m_{B'}}\cdot\overrightarrow{\sigma}\otimes I+I\otimes\overrightarrow{m_C}\cdot\overrightarrow{\sigma}+\sum_{i,j=1}^3s_{ij}^{BC}\sigma_i\otimes \sigma_j)$.

Substituting in Eq.(\ref{1.2}) and Eq.(\ref{1.3}) we obtain
\begin{equation} \label{2.4}
I=\sum_{k=1}^3(x_{0k}+x_{1k})\sum_{i=1}^3t_{ki}^{AB}(\sum_{j=1}^3m_{ij})\sum_{l=1}^3s^{BC}_{kl}(y_{0l}+y_{1l})
\end{equation}
and
\begin{equation}\label{2.5}
J=\sum_{k=1}^3(x_{0k}-x_{1k})\sum_{i=1}^3t_{ki}^{AB}(\sum_{j=1}^3n_{ij})\sum_{l=1}^3s^{BC}_{kl}(y_{0l}-y_{1l}).
\end{equation}

We can see that from Eq.(\ref{2.4}) and Eq.(\ref{2.5}) the value of $S$ depends largely on $t_{ij}^{AB}, s_{ij}^{BC}, \overrightarrow{x_i}, \overrightarrow{y_i}$ and $m_{ij}, n_{ij}$.
We seek to find the maximum value of the bilocality inequality (\ref{1.1}) for arbitrary pairs of mixed states using particle swarm optimization(PSO). According to PSO, we can find $m_{ij}, n_{ij}$ such that $S^{max}>2$ when $\rho_{AB}$ and $\rho_{BC}$ both are entangled states. This means exists two different entangled states violation of the bilocality inequality. Now let's think about an entangled state $\rho_{AB}$ distributes to Alice and Bob, and another separable state $\rho_{BC}$ distributes to Bob and Charles. In this case, can we get two states that violate the standard bilocality inequality (\ref{1.1})?

In particular, consider the case where Alice-Bob, as well as Bob-Charles, share a noisy Bell state, a so-called Werner state, the best-known class of mixed states. For qubits, the Werner state is given by
$$\rho=p|\phi^+\rangle\langle\phi^+|+(1-p)I/4,$$
where $|\phi^+\rangle=\frac{1}{\sqrt{2}}(|00\rangle+|11\rangle)$ is the singlet state, $I$ is the $4\times 4$ identity matrix and $p\in[0,1]$. The Werner state is entangled if and only if $p>1/3$. The Werner state is pure only when $p=1$.

We consider a specific quantum implementation of the bilocality experiment illustrated in Fig.1. The both sources emit pairs of qubits corresponding to Werner states. $\rho_{AB}=p|\phi^+\rangle\langle\phi^+|+(1-p)I/4$ and $\rho_{BC}=q|\phi^+\rangle\langle\phi^+|+(1-q)I/4$ (with $p,q\in[0,1]).$

Let's transform Werner state into something similar to Eq.(\ref{2.1}). We can get that
\begin{align}
\rho_{AB}=1/4(I\otimes I+p\sigma_1\otimes\sigma_1-p\sigma_2\otimes\sigma_2+p\sigma_3\otimes\sigma_3). \nonumber
\end{align}
Similarly, we express $\rho_{BC}$.

We can rewrite $I$ and $J$, substituting in Eq.(\ref{2.4}) and Eq.(\ref{2.5}), we obtain
\begin{equation}\begin{split}I=&[(x_{01}+x_{11})p(\sum_{i=1}^3m_{1i})-(x_{02}+x_{12})p(\sum_{i=1}^3m_{2i})+
(x_{03}+x_{13})p(\sum_{i=1}^3m_{3i})]\\
&\times[q(y_{01}+y_{11})-q(y_{02}+y_{12})+q(y_{03}+y_{13})] \\
=&pq[(x_{01}+x_{11})(\sum_{i=1}^3m_{1i})-(x_{02}+x_{12})(\sum_{i=1}^3m_{2i})+
(x_{03}+x_{13})(\sum_{i=1}^3m_{3i})]\\
&\times[(y_{01}+y_{11})-(y_{02}+y_{12})+(y_{03}+y_{13})].
\end{split}\end{equation}
$I$ can be expressed as $I=pqI'$, where
\begin{equation}\begin{split}
I'=&[(x_{01}+x_{11})(\sum_{i=1}^3m_{1i})-(x_{02}+x_{12})(\sum_{i=1}^3m_{2i})+
(x_{03}+x_{13})(\sum_{i=1}^3m_{3i})]\\
&\times[(y_{01}+y_{11})-(y_{02}+y_{12})+(y_{03}+y_{13})].
\end{split}\end{equation}
Similarly, we express $J=pqJ'$, where
\begin{equation}\begin{split}
J'=&[(x_{01}-x_{11})(\sum_{i=1}^3n_{1i})-(x_{02}-x_{12})(\sum_{i=1}^3n_{2i})+
(x_{03}-x_{13})(\sum_{i=1}^3n_{3i})]\\
&\times[(y_{01}-y_{11})-(y_{02}-y_{12})+(y_{03}-y_{13})].
\end{split}\end{equation}
So we can get
\begin{equation}
S=\sqrt{pq}(\sqrt{|I'|}+\sqrt{|J'|}).
\end{equation}

Denote $S'=\sqrt{|I'|}+\sqrt{|J'|}$, then our goal is to maximize $S'$ with the Bloch vector $\overrightarrow{x_0}, \overrightarrow{x_1}, \overrightarrow{y_0}, \overrightarrow{y_1},$ and $n_{ij}, m_{ij}$ (with $i,j\in\{1,2,3\}$). Next we use particle swarm optimization (PSO) algorithm as a approach to calculate $S'^{max}$.

\section{Experimental scheme and results}

Particle swarm optimization (PSO) algorithms \cite{RJ}
are outstandingly successful for non-convex optimization.
PSO is a 'collective intelligence' strategy from the field of
machine learning that learns via trial-and-error and performs as well as or better than simulated annealing and
genetic algorithms \cite{SBDP,JW,PA}. We have shown that PSO also delivers an autonomous approach to design an optimal measurement strategy. Here the optimal measurement means that the measurement can measure more quantum states with violation of the bilocality inequality.

The first step of the protocol is the initialization of a population: a set of lists ${\{\overrightarrow{x_0}, \overrightarrow{x_1}, \overrightarrow{y_0}, \overrightarrow{y_1}, m_{ij},n_{ij}\}}$ of measurement, $\forall i,j\in{\{1,2,3\}}$, corresponding the algorithm chromosomes, is randomly generated. The fitness is given by $S'$.

To search for $\mathfrak{\rho}_{opt}$, the PSO algorithm models a 'swarm' of $\sharp$  'particles' ${p^{(1)},p^{(2)},\ldots,p^{(\sharp)}}$ that move in the search space $\mathcal{P}_{N}$. In this paper, $\sharp$ take values 30. A particle's position $\rho^{(i)}\in\mathcal{P}_{N}$ represents a candidate policy for measurement $\varphi$, which is initially chosen at random. Furthermore, $p^{(i)}$ remembers
the best position, $\hat{\rho}^{(i)}$,  it has visited so far (including its current position). In addition, $p^{(i)}$ communicates with other particles in its neighborhood $\mathcal{N}^{(i)}\subseteq{\{1,2,\ldots,\sharp\}}$. We adopt the common approach to set each $\mathcal{N}^{(i)}$ in a pre-defined way regardless of the particles' positions by arranging them in a ring topology: for $p^{(i)}$, all particles with maximum distance $r$ on the ring are in $\mathcal{N}^{(i)}$. In
iteration $t$, the PSO algorithm updates the position of
all particles in a round-based manner as follows.

$(i)$ Each particle $p^{(i)}$ samples $\tilde{S'}(\rho^{(i)})$ of its current position with $K$ trial runs.

$(ii)$ $p^{(i)}$ re-samples $\tilde{S'}(\hat{\rho}^{(i)})$ of its personal-best policy $\hat{\rho}^{(i)}$, and the performance of $\hat{\rho}^{(i)}$ is taken to be the arithmetic mean $\bar{S'}(\hat{\rho}^{(i)})$ of all sharpness evaluations.

$(iii)$ Each $p^{(i)}$ update $\hat{\rho}^{(i)}$ if $\tilde{S'}(\hat{\rho}^{(i)})>\bar{S'}(\hat{\rho}^{(i)})$ and

$(iv)$ communicates $\hat{\rho}^{(i)}$ and $\bar{S'}(\hat{\rho}^{(i)})$ to  all members of $\mathcal{N}^{(i)}$.

$(v)$ Each particle $p^{(i)}$ determines the sharpest policy $\Lambda^{(i)}=max_{j\in\mathcal{N}^{(i)}}\hat{\rho}^{(i)}$
found so far by any one particle in $\mathcal{N}^{(i)}$ (including itself) and

$(vi)$ moves to
\begin{equation}\label{2.1..}
\rho^{(i)}\leftarrow\rho^{(i)}+\omega\delta^{(i)},\,\,\,\,
\delta^{(i)}\leftarrow\delta^{(i)}+\beta_1\xi_1(\hat{\rho}^{(i)}-\rho^{(i)})+\beta_2\xi_2(\Lambda^{(i)}-\rho^{(i)}).
\end{equation}

The arrows indicate that the right value is assigned to
the left variable. The damping factor $\omega$ assists convergence, and $\xi_1,\xi_2$ are uniformly-distributed random numbers from the interval $[0,1]$ that are re-generated each time Eq (\ref{2.1..}) is evaluated. The 'exploitation weight' $\beta_1$ parametrizes the attraction of a particle to its personal
best position $\hat{\rho}^{(i)}$, and the 'exploration weight' $\beta_2$ describes attraction to the best position $\Lambda^{(i)}$ in the neighborhood. To improve convergence, we bound each component of $\omega\delta^{(i)}$ by a maximum value of $\nu_{max}$. The userspecified parameters $\omega,\beta_1,\beta_2$ and $\nu_{max}$ determine the swarm's behavior.  Tests indicate that $\omega=0.8$, $\beta_1=0.5$, $\beta_2=0.5$, and $\nu_{max}=0.2$  result in the highest probability to find an optimal policy.

%The $K$ trial runs for assessing sharpness can be simulated or performed with a real world-experiment. For finite $K$, the sampled sharpness has statistical errors that can prevent the PSO algorithm from learning optimal solutions \cite{TDJ}. We reduce sharpness errors by averaging over multiple samples in step $(ii)$ \cite{JYA}.

After 500 iterations using PSO alogorithm, we get $S'^{max}=4.0642$.\\

\centerline{\includegraphics[width=3.5in]{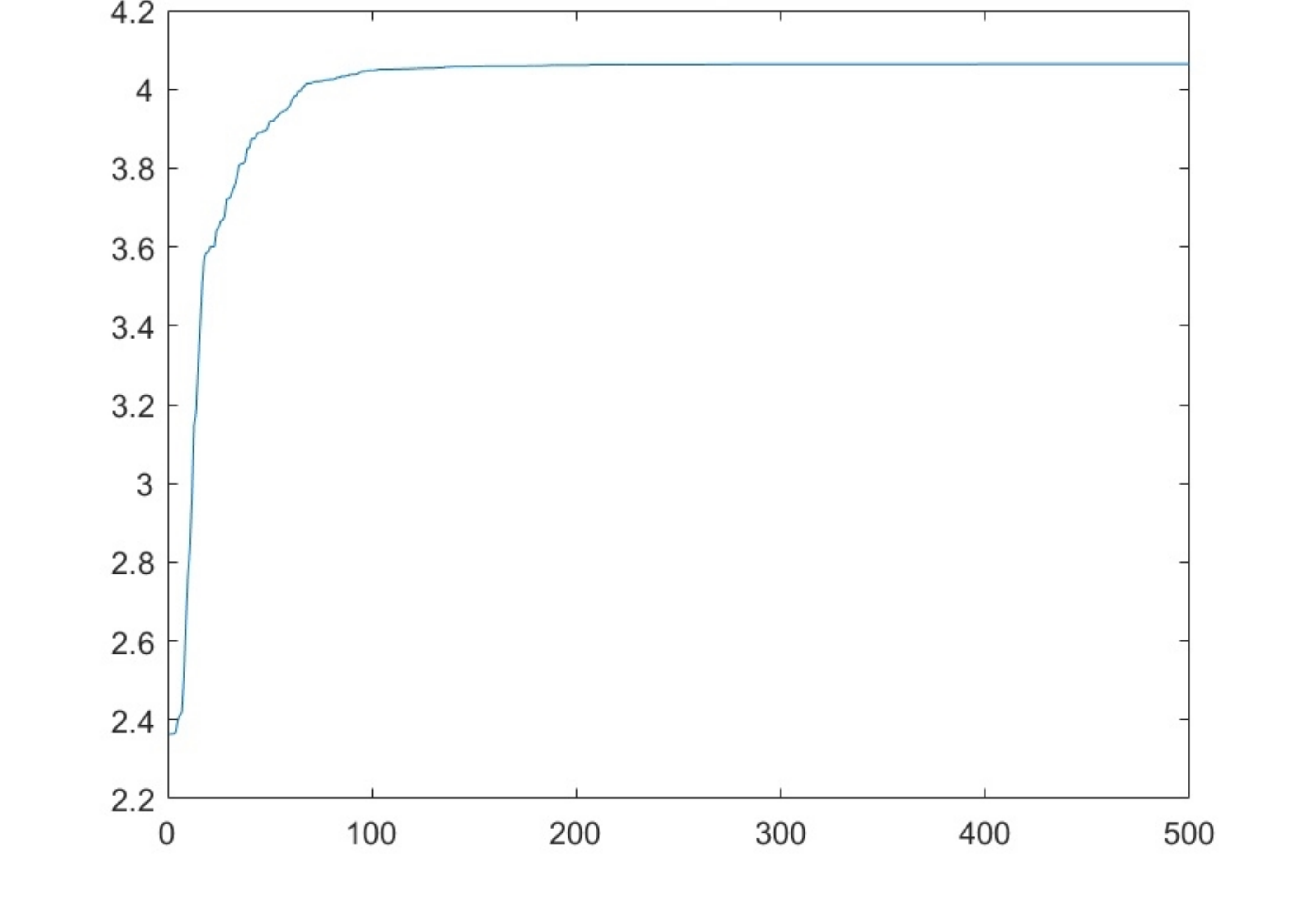}}
%\vspace*{6mm}

\vskip 0.5mm

%\centerline{\footnotesize Í¼1\ \ \  Gd$_{60}$Fe$_{30}$Al$_{10}$ Ìõ´øµÄ´Å»¯ÇúÏß \quad (origin µÈÈí¼þ»æÖÆµÄ²åÍ¼)}

%\vskip 1mm

\centerline{\parbox[c]{5cm}{\footnotesize Fig.~2.~Convergence of the $S'^{max}$}}

\vskip 0.55\baselineskip
%\begin{multicols}{2}

At this time, the corresponding measurements of Alice, Charles and Bob can be expressed as:
$$\overrightarrow{x_0}=(-0.9122,-0.1869,0.3647),$$
$$\overrightarrow{x_1}=(0.3321,-0.8910,0.3095),$$
$$\overrightarrow{y_0}=(-0.7915,-0.3305,0.5140),$$
$$\overrightarrow{y_1}=(-0.5737,0.5731,-0.5851),$$

$$
M=(m_{ij})=\left(
\begin{matrix}
   -0.1258  & -0.1882 & -0.2448 \\
   0.3078 & 0.4614 & 0.5996 \\
   0.1740 & 0.2606 & 0.3390
  \end{matrix}
  \right ),
$$

$$
N=(n_{ij})=\left(
\begin{matrix}
   -0.4062  & -0.5048 & -0.5051 \\
   -0.2797 & -0.3474 & -0.3476 \\
   -0.0049 & -0.0060 & -0.0060
  \end{matrix}
  \right ).
$$

In order to $S^{max}>2$, we consider the following expression:
\begin{align}
\sqrt{pq}(\sqrt{|I'|}+\sqrt{|J'|})>2. \\ \nonumber
\sqrt{pq}S'^{max}>2.    \nonumber
\end{align}
We thus get $pq\in[(\frac{2}{4.0642})^2,1]=[0.2422,1]$. This shows that we get a violation of our inequality (\ref{1.1}) for $pq\in[0.2422,1]$. The most interesting case is when $p=\frac{3.2}{4.1294}$ and $q=\frac{1}{3.1}$, as satisfied by the scope of the $pq$. In this case, $\rho_{AB}$ is an entangled state, $\rho_{BC}$ is a separable state and they are mixed states.

The result of this paper is to select the appropriate measurement to find the mixed Werner states which violate the bilocal inequality. Specifically, entangled state and separable state can violate the bilocality inequality (\ref{1.1}) if $p\in(\frac{1}{3},1]$ and $q\in[0,\frac{1}{3}]$.
Similarly, when $\rho_{BC}$ is an entangled state and $\rho_{AB}$ is a separable state, we can get $p$ and $q$, such that they violate the bilocality inequality. An evaluation of the relevant semidefinite program guarantees that a violation of bilocality is obtained whenever $p=q\geq83\%$
\cite{PAP}. However, we provide a better scope of $pq$ and find particular conclusion.


\begin{thebibliography}{99}

\bibitem{JS} J. S. Bell, On the einstein podolsky rosen paradox, \textit{Physics Physique Fizika}. 1(3),
195 (1964).
\bibitem{NDSV} N. Brunner, D. Cavalcanti, S. Pironio, V. Scarani and S. Wehner, Bell nonlocality, \textit{Rev. Mod. Phys}. 86, 419 (2014).
\bibitem{ANNSS} A. Acin, N. Brunner, N. Gisin, S. Massar, S. Pironio and V. Scarani, Device-independent security of
quantum cryptography against collective attacks, \textit{Phys. Rev. Lett}. 98, 230501 (2007).

\bibitem{MAH} M. Ben-Or, A. Hassidim and H. Pilpel, Quantum Multi Prover Interactive Proofs with Communicating
Provers, \textit{In: Proceedings of 49th Annual IEEE Symposium on Foundations of Computer Science, Los Alamitos, CA: IEEE}. (2008).

\bibitem{HRS} H. Buhrman, R. Cleve, S. Massar and R. de Wolf, Non-locality and Communication Complexity, \textit{Rev.
Mod. Phys}. 82, 665 (2010).

\bibitem{JHB} J. Bri$\ddot{\mathrm{e}}$t, H. Buhrman and B. Toner, A generalized Grothendieck inequality and entanglement in XOR
games, arXiv:0901.2009.

\bibitem{CNS1}C. Branciard, N. Gisin and S. Pironio, Characterizing the nonlocal correlations created via entanglement swapping, \textit{ Phys. Rev. Lett}. 104, 170401 (2010).
\bibitem{CDNS} C. Branciard, D. Rosset, N. Gisin and S. Pironio, Bilocal versus nonbilocal correlations in entanglement-swapping experiments, \textit{Phys. Rev. A}. 85, 032119 (2012).
\bibitem{TF} T. Fritz, Beyond Bell's theorem: correlation scenarios, \textit{ New J. Phys}. 14, 103001 (2012).
\bibitem{APDA} A. Tavakoli, P. Skrzypczyk, D. Cavalcanti and A. Ac¨ªn, Nonlocal correlations in the star-network configuration,
\textit{Phys. Rev. A}. 90, 062109 (2014).
\bibitem{RRJD} R. Chaves, R. Kueng, J. B. Brask and D. Gross, Unifying framework for relaxations of the causal assumptions in Bell's theorem, \textit{Phys. Rev. Lett}. 114, 140403 (2015).

\bibitem{CB} C. Brukner, Quantum causality, \textit{Nat. Phys}. 10, 259 (2014).
\bibitem{RCD} R. Chaves, C. Majenz and D. Gross, Information¨Ctheoretic implications of quantum causal structures, \textit{Nat. Commun}. 6, 5766
(2015).
\bibitem{FS} F. Costa and S. Shrapnel, Quantum causal modelling, \textit{New J. Phys}. 18, 063032 (2016).
\bibitem{TF2} T. Fritz, Beyond Bell's theorem II: Scenarios with arbitrary causal structure, \textit{Commun. Math. Phys}. 341, 391 (2016).
\bibitem{ML} M-X. Luo, Computationally efficient nonlinear bell inequalities for quantum networks, \textit{Phys. Rev. Lett}. 120, 140402 (2018).
\bibitem{ERT} E. Wolfe, R. W. Spekkens and T. Fritz, The Inflation Technique for Causal Inference with Latent Variables, \textit{J. Causal Inference}. 7, 2 (2019).
\bibitem{EAMDAM} E. Wolfe, A. Pozas-Kerstjens, M. Grinberg, D. Rosset, A. Ac¨ªn and M. Navascues, Quantum Inflation: A General Approach to Quantum Causal Compatibility, arXiv:1909.10519.
\bibitem{MESNNS} M. O. Renou, E. Baumer, S. Boreiri, N. Brunner, N. Gisin and S. Beigi, Genuine quantum nonlocality in the triangle network, \textit{Phys. Rev. Lett}. 123, 140401 (2019).
\bibitem{MYSSNN} M-O. Renou, Y. Wang, S. Boreiri, S. Beigi, N. Gisin and N. Brunner, Limits on Correlations in Networks for Quantum and No-Signaling Resources, \textit{Phys. Rev. Lett}. 123, 070403 (2019).
\bibitem{NJYAESN} N. Gisin, J. D. Bancal, Y. Cai, A. Tavakoli, E. Z. Cruzeiro, S. Popescu and N. Brunner, Constraints on nonlocality in networks from no-signaling and independence, \textit{Nat Commun}. 11, 2378 (2020).
\bibitem{NDSVS} N. Brunner, D. Cavalcanti, S. Pironio, V. Scarani and S. Wehner, Publisher's note: Bell nonlocality, \textit{Rev. Mod. Phys}. 86, 419 (2014).
\bibitem{GMQTA} N. Gisin, Q. X. Mei, A. Tavakoli, M. O. Renou and N. Brunner, All entangled pure quantum states violate the bilocality inequality, \textit{Phys. Rev. A}. 96(2), 020304 (2017).
\bibitem{RPM} R. Horodecki, P. Horodecki and M. Horodecki, Violating Bell inequality by mixed spin-12 states: necessary and sufficient condition, \textit{Phys. Lett. A}.
200(5), 340 (1995).

\bibitem{RJ} R. Eberhart and J. Kennedy, A new optimizer using particle swarm theory, \textit{Proceedings of the Sixth International Symposium on Micro Machine and Human Science. IEEE}. (1995).



\bibitem{SBDP} S. Ethni, B. Zahawi, D. Giaouris and P. Acarnley, Comparison of Particle Swarm and Simulated Annealing Algorithms for Induction Motor Fault Identification,  \textit{2009 7th IEEE International Conference on Industrial Informatics, IEEE.} (2009).
\bibitem{JW} J. Kennedy and W. M. Spears, Matching Algorithms to Problems: An Experimental Test of the Particle Swarm and Some Genetic Algorithms on the Multimodal Problem Generator, \textit{1998 IEEE International Conference on Evolutionary Computation Proceedings. IEEE World Congress on Computational Intelligence (Cat. No. 98TH8360), IEEE.} (1998).
\bibitem{PA} P. Fourie and A. Groenwold, The particle swarm optimization algorithm in size and shape optimization, \textit{Struct. Multidiscipl. Optim}. 23(4), 259 (2002).


\bibitem{PAP} A. Pozas-Kerstjens (private communication).


\end{thebibliography}
\end{document}